\documentclass[prl,showpacs,twocolumn,amsfonts]{revtex4}

\usepackage{graphicx}
\usepackage{bm}

\begin{document}

\title{Stretching semiflexible filaments with quenched disorder}
\author{Panayotis Benetatos and Eugene M. Terentjev}
\affiliation{Cavendish Laboratory, University of Cambridge, J. J.
  Thomson Avenue, Cambridge, CB3 0HE, United Kingdom }

\date{\today}

\begin{abstract}
We study the effect of quenched randomness in the
arc-length dependent spontaneous curvature of a wormlike chain under
tension. In the
weakly bending approximation in two dimensions, we obtain analytic results for the
force-elongation curve and the width of transverse fluctuations. We
compare quenched and annealed disorder and conclude that the former
cannot always be reduced to a simple change in the stiffness of the
pure system. We also discuss the effect of a random transverse force
on the stretching response of a wormlike chain without spontaneous curvature.

\end{abstract}
\maketitle

{\it Introduction.---} The wormlike chain (WLC) model of semiflexible polymers treats the
macromolecules as one-dimensional locally inextensible curves with bending
rigidity \cite{KP,Saito,SS1,SS2,SS3}. Despite its simplicity and its initial success with
the experiments on stretching of ds-DNA \cite{Bustamante}, the
inherent complexity in the
microscopic architecture of biological filaments invites the
search for more realistic models.  One key feature of
many biopolymers, such as DNA or denatured proteins, which goes
beyond the classical WLC is their heterogeneity.  Their local architecture (e.g., spontaneous
curvature) \cite{Anselmi}  or their local elasticity (e.g., their bending rigidity) 
vary along the polymer contour. At large scales, this heterogeneity
may statistically behave as a random variable. It may be fixed in time such as that related to the
different base pair sequences in DNA. This randomness is treated as
quenched disorder. If the random inhomogeneity undergoes thermal
fluctuations, it is treated as annealed disorder. An example of the
latter is the reversible binding of proteins to DNA capable of
inducing spontaneous (intrinsic) curvature which has been studied in \cite{Rabin_Rap}.

The first studies of the effect of random base pair sequences on the
macroscopic statistical properties of DNA focused on the related random
bends (spontaneous curvature) and treated them on equal footing as the thermally excited bends,
thus considering the annealed disorder case
\cite{Trifonov,Schellman,SchellmanHarvey}. Bensimon {\it et al.}
considered the effect of random angles in the Kartky-Porod chain and
using numerical transfer matrix and Monte-Carlo methods found that, as far as
stretching is concerned, the disordered chain behaves as a homogeneous
chain with a renormalized persistence length
\cite{Bensimon}. P. Nelson calculated the effect of weak disorder on
the entropic elasticity of a flexible rod with random kinks and
found that the sole effect was a reduction in the apparent bending
stiffness with the twist stiffness remaining unchanged
\cite{PhilNel}. A different way of modeling disorder has been followed
by Debnath and Cherayil \cite{Debnath} who considered the chain under
tension as consisting of random A-B blocks with different elastic
constants. The blocks were treated as ``Gaussian'' semiflexible polymers
(without the local inextensibility constraint). 
Muhuri and Rao
\cite{Rao} studied finite size effects in the Kratky-Porod chain with
a random sequence of stiffness constants. 

In a recent paper \cite{paper1}, we studied the response of a weakly
bending WLC with arc-length dependent spontaneous curvature to a stretching force applied
at its ends. We specifically considered the case of sinusoidally
varying spontaneous curvature which allows us to treat the general
case via Fourier transformation. This simple and analytically
tractable model appears to be particularly well suited for the study
of quenched disorder in the spontaneous curvature of a filament under
tension. This is the subject of the present paper. A similar
calculation with a different method (which is in principle valid) was attempted in Ref. \cite{Zhou2}, but a mistake
in the way the thermodynamic limit was taken led to the incorrect conclusion that
sequence disorder has no effect on elasticity. Here, we use the
replica trick to calculate the effect of uncorrelated quenched disorder and we
obtain the force-extension relationsip as well as
force-transverse-width relationship. We compare our results with those
obtained in the context of annealed disorder. Although, as we show,
the two cases (quenched and annealed) are indistinguishable in the
limit of weak disorder, our results for quenched disorder hold 
for arbitrary strength.

We also study the adaptation to the WLC of the classical random-force
model which Larkin had introduced in the context of the Abrikosov lattice
\cite{Larkin}. As with quenched disorder in the spontaneous curvature
in the weakly bending approximation, the coupling of a random force is
linear and this makes its analytical
treatment straightforward. The effect of quenched random forces on the
WLC conformations has recently been analyzed in
\cite{Marcel}. Here, we deal with their effect on the stretching response
in parallel with our study of the random spontaneous curvature.


{\it Model.---} We consider a WLC fluctuating in two dimensions under a
strong stretching force applied at its end-points. 
The restriction for
the chain to remain confined to a single plane is made to allow
concise analytical treatment. It certainly applies to stretching
experiments on a surface \cite{Seifert}. Although it is known the
difference between $2d$ and $3d$, when the curvature of a freely bending
WLC is concerned, to be
non-trivial \cite{Rabin3d}, here we stay within the weakly
bending approximation where it can be shown, using the Monge
gauge \cite{PBdepin}, that the two transverse directions of a
$(1+2)$-dimensional chain decouple. Our recovery, for a special case
of disorder,
of the main result of Ref. \cite{PhilNel} which was based on a $3d$
model further
attests to the validity of the main features of our results beyond $2d$. 
Quenched disorder
along the polymer contour is represented by a random distribution of the
arc-length dependent spontaneous curvature. The strong stretching
force allows us to use the weakly bending approximation which
significantly simplifies the analytical treatment of the filament
response as shown in Ref. \cite{paper1}. The WLC conformations are
parametrized by the displacement $y(s)$, transverse to the direction
of the pulling force ($x$), as a function of the arc-length position $s$. The elastic energy functional is given by
\begin{eqnarray}
\label{H_c}
{\cal H}_c[y(s)]&= &\frac{\kappa}{2}\int_0^L ds \Big[\Big(\frac{\partial^2
  y(s)}{\partial s^2}\Big)-{\tilde c}(s)\Big]^2\nonumber\\
& &+\frac{1}{2}f\int_0^L
  ds \Big(\frac{\partial
  y(s)}{\partial s}\Big)^2 -fL\;,
\end{eqnarray}
where $\kappa$ is the bending rigidity related to the persistence length
via $\kappa=\frac{1}{2}L_p k_B T$, $f$ is the applied stretching force, $L$ is
the total contour length, and ${\tilde c}(s)$ is the signed
spontatneous curvature. The probability distribution of  ${\tilde
  c}(s)$  is given by 
\begin{eqnarray}
\label{Pofc}
{\cal P}[{\tilde c}(s)] \sim \exp\Big[-\int_0^L ds
\frac{1}{2\Delta}_c \big({\tilde c}(s)\big)^2\Big]\;,
\end{eqnarray}
and is completely characterized by $\overline{{\tilde c}(s)}=0$ and
$\overline{{\tilde c}(s){\tilde c}(s')}={\Delta}_c{\delta}(s-s')$, where
the overbar denotes an average over the disorder. The delta function
reflects the assumption of uncorrelated disorder along the polymer backbone.

Experimentally observable quantities of long polymers with quenched
disorder are expected to be self-averaging
and are calculated by averaging the free energy over the distribution
of disorder \cite{Dotsenko}. This requires averaging the logarithm of the partition
function, a goal which is achieved by employing the standard replica
trick \cite{EA}. Averaging the partition function itself over disorder corresponds to
the annealed system which has been studied in Ref. \cite{Zhou2} where
it is called the ``disorder-first system.''

Besides disorder which is related to the spontaneous curvature, in this paper, we also consider the effect of a random transverse force on
the force-extension relationship of a WLC without
spontaneous curvature. The elastic energy functional of such a system is
given by
\begin{eqnarray}
\label{H_g}
{\cal H}_g[y(s)]&= &\frac{\kappa}{2}\int_0^L ds \Big(\frac{\partial^2
  y(s)}{\partial s^2}\Big)^2 +  \int_0^L ds g(s)y(s)\nonumber\\
& &+\frac{1}{2}f\int_0^L
  ds \Big(\frac{\partial
  y(s)}{\partial s}\Big)^2 -fL\;.
\end{eqnarray}
As usual, the distribution of the random force $g(s)$ is assumed to be
completely determined by $\overline{g(s)}=0$ and
$\overline{g(s)g(s')}={\Delta}_g{\delta}(s-s')$. The random coupling
term in Eq. (\ref{H_g}) can be viewed as expressing the
  interaction of a stretched semiflexible polyampholyte with a
  quenched linear charge density $g(s)/E$ and a uniform transverse
  field $E$. An alternative interpretation of Eq. (\ref{H_g}) views
  $g(s)$ as representing a random
  interaction with the crowded environment in which the
  filament fluctuates. Strictly speaking, quenched disorder in the
  embedding medium (environment) should entail a term $\int_0^L ds g(x(s))y(s)$, where $x(s)=s-\frac{1}{2}\int_0^s
  ds' (\partial_{s'} y(s'))^2$, instead of the one written above, but $g(s)$ is expected to capture the
  leading-order behavior in the weakly bending 
  limit.


{\it Replica Trick.---} The standard method to deal with the quenched disorder average
involves averaging over $n$ identical non-interacting copies (replicas) of the
system \cite{EA}. In the end, the replica limit $n\rightarrow0$ is taken. The free
energy is calculated via $\overline{\ln
  Z}=\lim_{n \rightarrow 0}(\overline{ Z^n}-1)/n$, where $Z$
is the partition function of the system with a certain realization of
the disorder. $\overline{Z^n}$ can be expressed as
$\overline{Z^n}=\exp(-{\cal H}^{(rep)}/k_BT)$. 

For the random-spontaneous-curvature system described by
Eq. (\ref{H_c}), the replica ``Hamiltonian'' takes the form 
\begin{eqnarray}
\label{H_crep}
&{\cal H}_c^{(rep)}&=\frac{1}{2}\int_0^L ds
\displaystyle\sum_{a=1}^n \Big[\kappa \Big(\frac{\partial^2
  y_a(s)}{\partial s^2}\Big)^2 +f\Big(\frac{\partial
  y_a(s)}{\partial s}\Big)^2 \Big]\nonumber\\
&&- \frac{{\Delta}_c {\kappa}^2}{2 k_B T(1+n\kappa{\Delta}_c/k_BT)} \int_0^L ds \displaystyle\sum_{a,b=1}^n\Big(\frac{\partial^2
  y_a(s)}{\partial s^2}\Big)\Big(\frac{\partial^2
  y_b(s)}{\partial s^2}\Big),
\end{eqnarray}
where we have omitted constant terms and the subscripts $a,\;
 b$ label
the replicas of the original system. 
In Ref. \cite{paper1}, we have shown that in the strong stretching
regime, where $f \gg \kappa/L^2$, the details of the boundary
conditions become irrelevant. Therefore we can assume hinged-hinged
boundary conditions and a chain with vanishing curvature at its
end-points. In this case, we can use the Fourier decomposition 
\begin{eqnarray}
\label{Fourier}
y_a(s)=\displaystyle\sum_{m=1}^{\infty}A_a^{(m)}\sin(q_ms)\;,
\end{eqnarray}
where $q_m=\pi m/L$ is the wavenumber of the corresponding $m$-mode,
and express the replica ``Hamiltionian'' of
Eq. (\ref{H_crep}) in a quadratic matrix form. 
Correlators can be calculated using the
Sherman-Morrison formula from linear algebra. In the replica limit ($n\rightarrow 0$), we get
\begin{eqnarray}
\label{ccorr}
\langle A_a^{(m)}A_b^{(m)}\rangle=\frac{2k_BT}{Lq_m^{2}(\kappa q_m^2
  +f)}\delta_{ab}+\frac{2 \Delta_c \kappa^2 }{L(\kappa q_m^2 +f)^2}{\bf 1}_{ab}\;,
\end{eqnarray}
where ${\bf 1}$ is an $n\times n$ matrix with all of its elements equal to
1. Using the matrix determinant lemma and taking the replica limit we also calculate
the disorder averaged free energy and obtain
\begin{eqnarray}
\label{Frenc}
\overline{G}_c&= & -\frac{k_B T}{2}
\displaystyle\sum_{m=1}^{\infty}\Big[\ln \Big(\frac{\pi}{k_BT}(\kappa
q_m^2 + f)\Big)\nonumber\\
& & - \frac{1}{k_B T}\frac{\kappa^2 \Delta_c q_m^2}{(\kappa
  q_m^2 +f)}\Big]\;. 
\end{eqnarray}

For the random-force system described by Eq. (\ref{H_g}), the
corresponding replica
``Hamiltonian'' is given by the analoguous procedure:
\begin{eqnarray}
{\cal H}_g^{(rep)}&= &\frac{1}{2}\int_0^L ds
\displaystyle\sum_{a=1}^n \Big[\kappa \Big(\frac{\partial^2
  y_a(s)}{\partial s^2}\Big)^2 + f \Big(\frac{\partial
  y_a(s)}{\partial s}\Big)^2\Big] \nonumber\\
& &- \frac{{\Delta}_g}{2 k_B T} \int_0^L ds
\displaystyle\sum_{a,b=1}^n y_a(s)y_b(s)\;.
\end{eqnarray}
Using similar calculations as in the case of random spontaneous
curvature, we obtain the correlators
\begin{eqnarray}
\label{gcorr}
\langle A_a^{(m)}A_b^{(m)}\rangle=\frac{2k_BT}{Lq_m^2(\kappa q_m^2
  +f)}\delta_{ab}+\frac{2 \Delta_g }{Lq_m^4(\kappa q_m^2 +f)^2}{\bf 1}_{ab}\;,
\end{eqnarray}
and the disorder averaged free energy
\begin{eqnarray}
\label{Freng}
\overline{G}_g&= &\frac{k_B T}{2}
\displaystyle\sum_{m=1}^{\infty}\Big[\ln \Big(\frac{\pi}{k_B T}(\kappa
q_m^2 + f)\Big)\nonumber\\
& &-\frac{1}{k_B T}\frac{
\Delta_g }{q_m^2(\kappa
  q_m^2 +f)}\Big]\;. 
\end{eqnarray}
 

{\it  Force-Extension Relationship.---} The average projected length of the filament in the
direction of the stretching force is given by
\begin{eqnarray}
\label{ave(x)}
\overline{\langle x(L) \rangle} = L-\frac{1}{2}\int_0^L ds
\overline{\Big \langle \Big(\frac{\partial y(s)}{\partial s}\Big)^2
  \Big \rangle}\;,
\end{eqnarray}
where
\begin{eqnarray}
\overline{\Big\langle \Big(\frac{\partial y(s)}{\partial
    s}\Big)^2 \Big\rangle}=\frac{1}{2}\displaystyle\sum_{m=1}^{\infty}
q_m^2 \overline{\langle (A^{(m)})^2\rangle}\;
\end{eqnarray}
and 
\begin{eqnarray}
\overline{\langle (A^{(m)})^2\rangle}=\displaystyle\lim_{n\rightarrow
  0}\frac{1}{n}\displaystyle\sum_{a=1}^n{\langle A_a^{(m)} A_a^{(m)} \rangle}\;.
\end{eqnarray}
In the strong stretching regime, defined by $f\gg f_{cr} \gtrsim f_L$,
where $f_{cr}\equiv \kappa/L_p^2$ and $f_L\equiv \kappa /L^2$ 
(the shorthand notation defining the characteristic force scales)
, we obtain
\begin{eqnarray}
\label{fexc}
\frac{\overline{\langle
    x(L)\rangle}}{L}-1=-\frac{1}{2}\Big(\frac{f_{cr}}{f}\Big)^{1/2}-\frac{1}{8}\Delta_c
L\Big(\frac{f_{L}}{f}\Big)^{1/2}\;.
\end{eqnarray}

Equation (\ref{fexc}) is the central result of our paper. The first
term in the right-hand side (rhs) is the well-known expression associated with the ironing out of the thermal
undulations \cite{MS}. The second term comes from straightening the quenched
random
undulations related to the spontaneous curvature. This result implies that, as
far as the force-extension response is concerned, uncorrelated
quenched disorder in the spontaneous curvature acts as an effective
linear increase in the temperature given by $k_B \delta T=\kappa
\Delta_c/2$ (using $L_p=2\kappa/k_BT$). 
We also
point out that for given bending rigidity and disorder strength, the
effect of disorder on the force-extension relationship is independent
of the polymer size $L$. This result holds for {\it arbitrary} strength of
disorder, provided that the strong stretching and weakly bending
assumptions are fulfilled. Our result does not contradict the claim by
Marko and Siggia \cite{MS} that disorder related to intrinsic bends with large
radius of curvature compared to the persistence length do not alter
the large $f$ limit. In fact, {\it any} arc-length dependent spontanous
curvature whose Fourier spectrum has bounded amplitude does not alter
this limit. As we have shown in \cite{paper1}, the response of a
weakly bending WLC with arbitrary (but bounded) spontaneous curvature is
the superposition of the responses corresponding to sinusoidally
varying modes of curvature. The latter contribute terms which scale
as $\sim f^{-2}$ and become negligible compared with the thermal term
$\sim f^{-1/2}$ as the stretching force increases. The crucial
difference with the case of uncorrelated disorder is that the 
latter contains bends of unbounded sharpness in a similar fashion as the
thermal excitations do. That is the physical reason why the disorder term in
Eq. (\ref{fexc}) has the same form as the thermal one. Of course,
uncorrelated disorder with zero correlation length is a mathematical
abstraction. In real systems, there will be a maximum wavenumber,
$q_{\rm max}$, in the spectrum of random undulations of the spontaneous
curvature. For $f\ll \kappa q_{\rm max}^2$ \cite{paper1}, disorder
approximately behaves as
uncorrelated and our Eq. (\ref{fexc}) is expected to hold. For $f\gg
\kappa q_{\rm max}^2$, the Marko-Siggia (thermal) limit for large $f$ will prevail.

An alternative way to obtain the force-extension relationship is by
taking the derivative of the disorder-averaged free energy given in
Eq. (\ref{Frenc}) with respect to the stretching force:
$\partial_f \overline{G}_c= (\overline{\langle
    x(L)\rangle}-L)/L\;$.
As expected, both ways yield exactly the same result.

In Ref. \cite{Zhou2}, it is shown that the effect of annealed disorder
in the spontaneous curvature amounts to the replacement of the
original bending rigidity $\kappa$ by

$\kappa_{eff}=\kappa /(1+\Delta_c \kappa / k_BT)\;$.
This implies that for
{\it weak} disorder, defined by $\Delta_c \ll k_BT/\kappa$, the
approximate linear relation holds
$\kappa_{eff}\approx \kappa (1-\Delta_c \kappa/k_BT)$
and we immediately recover the
response described by Eq. (\ref{fexc}). In this regime, quenched
disorder has the same effect as annealed disorder. This is the regime
treated in Ref. \cite{PhilNel}. The difference
between the two types of disorder becomes significant in the case of
strong disorder, where $\kappa_{eff}\approx k_BT/\Delta_c$. In the
quenched case, the second term in the rhs of Eq. (\ref{fexc}) dominates
the response. In the annealed case, we would have gotten $-\frac{1}{4}\Delta_c L
(f_L/f)^{1/2}$ instead which differs by a factor of $2$. A more
significant qualitative difference between quenched and annealed
disorder can be deduced from the form of Eq. (\ref{ccorr}). Except for
the weak regime, the effect of quenched disorder cannot be reduced to a
simple renormalization of the bending rigidity.

The force-extension relationship for the random-force system of
Eq. (\ref{H_g}) can be calculated in a similar fashion, taking into
account Eqs. (\ref{gcorr}) or (\ref{Freng}). For strong stretching, $f\gg
f_{cr} \gtrsim f_L$, we obtain
\begin{eqnarray}
\label{fexg}
\frac{\overline{\langle
    x(L)\rangle}}{L}-1=-\frac{1}{2}\Big(\frac{f_{cr}}{f}\Big)^{1/2}-\frac{1}{6}\frac{\Delta_g
L}{f^2}\;.
\end{eqnarray}
We notice that as the stretching force increases, it quickly irons out the
bends caused by the random force and the response is dominated by the
classical 
thermal undulations. On the other hand, for given stretching force,
bending rigidity, temperature, and disorder strength, the effect of
disorder grows linearly with the contour length $L$ and becomes
dominant in the thermodynamic limit ($f$ needs to increase accordingly
in order to stay within the weakly bending approximation). This is an interesting similarity
with the random-force-induced destruction of long-range order in the
Abrikosov lattice according to the Larkin model
\cite{Larkin}. Comparing Eq. (\ref{fexg}) with the result obtained in
Ref. \cite{paper1} for the force-extension relationship of a WLC with sinusoidally varying spontaneous curvature along the
polymer contour, we see that the random force effectively acts as
spontaneous curvature of that type with amplitude 
$ c_{eff}=(\Delta_g L/3 \kappa^2)^{1/4}\;$.

{\it Transverse Fluctuations.---} The shape of transverse fluctuations of a stretched filament,
$\overline {\langle (y(s))^2 \rangle}$, is a useful diagnostic tool of
its elasticity distinct from its extension in the direction of the
pulling force, $\overline {\langle x(L) \rangle}$
\cite{upenn_eng}. Using the correlators of Eq. (\ref{ccorr}), we
calculate the width of the transverse fluctuations at the mid-point
($s=L/2$), for $f\gg f_L$, and obtain
\begin{eqnarray}
\label{transvc}
\overline{\Big\langle \Big(y(s=\frac{L}{2})\Big)^2 \Big\rangle}=\frac{L k_B
  T}{4}\frac{1}{f}+\frac{\Delta_c \kappa^{3/2}}{4}\frac{1}{f^{3/2}}\;.
\end{eqnarray}
For strong disorder, the second term in the rhs of the above equation,
which scales as $\sim f^{-3/2}$, can dominate over a range of forces before it is overtaken by the thermal
term which scales as $\sim f^{-1}$. In the annealed case the
disorder-related term is absent, whereas in the case of weak disorder,
$\Delta_c \ll k_B T/\kappa$, it is negligible. We point out that strong quenched disorder in the spontaneous
curvature causes a {\it qualitatively} different behavior in the
response of transverse fluctuations which cannot be simply reduced to
a renormalization of the persistence length of the undisordered
wormlike chain. We see that the claim by Bensimon {\it et al.}
\cite{Bensimon} that ``a random Kratky-Porod chain with a certain type
of disorder is well approximated by a pure chain with an effective
elastic constant'' is not totally correct. It only applies to the
force-elongation curve and cannot be extended to the transverse
fluctuations of a strongly disordered WLC. 

For the random-force system, the width of the transverse fluctuations at the mid-point
($s=L/2$), for $f\gg f_L$, is given by 
\begin{eqnarray}
\label{transvc}
\overline{\Big\langle \Big(y(s=\frac{L}{2})\Big)^2 \Big\rangle}=\frac{L k_B
  T}{4}\frac{1}{f}+\frac{\Delta_g L^3}{48}\frac{1}{f^{2}}\;.
\end{eqnarray}
The random-force contribution scales as a higher power of the filament
length $L$ and thus is relevant in the the long chain limit.

{\it Conclusions.---} We studied how the uncorrelated disorder in the spontaneous
curvature of a weakly bending WLC affects its response under strong tension. The
force-elongation curve for quenched disorder is identical to that of a
pure system at a higher temperature. The quenched case is identical to
the annealed case in the limit of weak disorder where its effect can
be reduced to a decrease in the bending stiffness. This is no longer
true as the disorder becomes stronger and our results hold for
arbitrary strength. The effect of strong quenched disorder on the
force dependence of the width of transverse fluctuations cannot be
simply reduced to an effective change in the temperature or the
bending stiffness of the pure chain. 

We also studied the influence of uncorrelated quenched disorder
in the transverse force along the contour of a WLC without
spontaneous curvature under
tension. In this case, the force-elongation curve is
identical to that of a pure WLC with sinusoidal spontaneous
curvature. We also find that the effect of the random force increases
with the length of the chain. A strong random force may qualitatively
alter the force dependence of the transverse fluctuations over a range
of pulling forces.

This work was supported by
EPSRC via the TCM Programme Grant. P.B. acknowledges discussions with
M. Hennes and K. Kroy.


\begin{thebibliography}{}





\bibitem{KP}
O. Kratky and G. Porod, Recl. Trav. Chim. Pays-Bas {\bf 68}, 1106 (1949).

\bibitem{Saito}
N. Sait\^o, K. Takahashi and Y. Yunoki, J. Phys. Soc. Jpn. {\bf 22}, 219 (1967).

\bibitem{SS1}
J. Samuel and S. Sinha, Phys. Rev. E {\bf 66}, 050801 (2002).

\bibitem{SS2}
A. Ghosh, J. Samuel, and S. Sinha,  Phys. Rev. E {\bf 76}, 061801 (2007).

\bibitem{SS3}
S. Sinha and J. Samuel,  Phys. Rev. E {\bf 71}, 021104 (2005).

\bibitem{Bustamante}
C. Bustamante, S. B. Smith, J. Liphardt, and D. Smith,
Curr. Opin. Struct. Biol. {\bf 10}, 279 (2000).

\bibitem{Anselmi}
C. Anselmi, P. De Santis, R. Paparcone, M. Savino, and A. Scipioni,
Biophys. Chem. {\bf 95}, 23 (2002).

\bibitem{Rabin_Rap}
S. M. Rappaport and Y. Rabin, Phys. Rev. Lett.  {\bf 101}, 038101 (2008).


\bibitem{Trifonov}
E. N. Trifonov, R. K.-Z. Tan, and S. C. Harvey, in {\it  DNA Bending and Curvature}, edited by W. K. Olson, M. H. Sarma, and M. Sundaralingam (Adenine Press, Schenectady, 1987).

\bibitem{Schellman}
J. A. Schellman, Biophys. Chem {\bf 11}, 329 (1980)

\bibitem{SchellmanHarvey}
J. A. Schellman and S. C. Harvey, Biophys. Chem. {\bf 55}, 95 (1995).


\bibitem{Bensimon}
D. Bensimon, D. Dohmi, and M. Mezard, Europhys. Lett. {\bf 42}, 97 (1998).


\bibitem{PhilNel}
P. C. Nelson, Phys. Rev. Lett. {\bf 80}, 5810 (1998).


\bibitem{Debnath}
P. Debnath and B. J. Cherayil, J. Chem. Phys. {\bf 118}, 1970 (2003).


\bibitem{Rao}
S. Muhuri and M. Rao, J. Stat. Mech.: Theory Exp. P02005 (2010).




\bibitem{paper1}
P. Benetatos and E. M. Terentjev,  Phys. Rev. E {\bf 81}, 031802 (2010).

\bibitem{Zhou2}
Z. Zhou and B. Jo\'os, Phys. Rev. E {\bf 80}, 061911 (2009). 


\bibitem{Larkin}
A. I. Larkin, Sov. Phys. JETP {\bf 31}, 784 (1970).


\bibitem{Marcel}
M. Hennes and K. Kroy, unpublished.


\bibitem{Seifert}
B. Meier, U. Seifert, and J. O. R\"adler, Europhys. Lett. {\bf 60},
622 (2002).

\bibitem{Rabin3d}
S. M. Rappaport and Y. Rabin, Phys. Rev E {\bf 80}, 052801 (2009).


\bibitem{PBdepin}
P. Benetatos and E. Frey, Phys. Rev. E {\bf 67}, 051108 (2003).

\bibitem{Dotsenko}
V. Dotsenko, {\it Introduction to the Replica Theory of Disordered
  Statistical Systems} (CUP, Cambridge, 2001).


\bibitem{EA}
S. F. Edwards and P. W. Anderson, J. Phys. F: Metal. Phys. {\bf 5},
965 (1975).

\bibitem{MS}
J. F. Marko and E. D. Siggia, Macromolecules {\bf 28}, 8759 (1995).



\bibitem{upenn_eng}
P. K. Purohit, M. E. Arsenault, Y. Goldman and H. H. Bau,
Int. J. Non Linear Mech. {\bf 43}, 1056 (2008).







\end{thebibliography}
\end{document}